\documentclass[a4paper,UKenglish,cleveref, autoref, thm-restate]{lipics-v2021}

\hideLIPIcs  

\title{Tight Cycle-Cover Inequalities and Improved Approximation for Shortest Common Superstring with Reverse Complements} 

\titlerunning{Tight Cycle-Cover Inequalities and Improved Approximation for SCS-RC} 

\author{Ryosuke Yamano}{Department of Computer Science, Graduate School of Information Science and Technology, The University of Tokyo, Japan \and Division of Medical Data Informatics, Human Genome Center, Institute of Medical Science, The University of Tokyo, Japan}{ryoyamano15@g.ecc.u-tokyo.ac.jp}{https://orcid.org/0009-0002-1683-5179}{}
\author{Tetsuo Shibuya}{Division of Medical Data Informatics, Human Genome Center, Institute of Medical Science, The University of Tokyo, Japan}
{tshibuya@hgc.jp}{https://orcid.org/0000-0003-1514-5766}{}

\authorrunning{R. Yamano and T. Shibuya} 

\Copyright{Ryosuke Yamano and Tetsuo Shibuya} 



\category{} 

\relatedversion{} 

\funding{This work was supported by MEXT KAKENHI Grant Number 23K28035.}

\nolinenumbers 

\usepackage[ruled,vlined]{algorithm2e}

\EventEditors{John Q. Open and Joan R. Access}
\EventNoEds{2}
\EventLongTitle{42nd Conference on Very Important Topics (CVIT 2016)}
\EventShortTitle{CVIT 2016}
\EventAcronym{CVIT}
\EventYear{2016}
\EventDate{December 24--27, 2016}
\EventLocation{Little Whinging, United Kingdom}
\EventLogo{}
\SeriesVolume{42}
\ArticleNo{23}

\begin{document}

\maketitle

\begin{abstract}
The Shortest Common Superstring (SCS) problem is a classical
idealized model of genome assembly.
However, SCS does not account for the double-stranded nature
of DNA: sequencing reads may originate from either strand
without a known orientation.
This motivates a natural variant, Shortest Common Superstring
with Reverse Complements (SCS-RC), which asks for a shortest
string containing each input string or its reverse complement
as a substring.
For SCS, the \textsf{MGREEDY} algorithm constructs
an optimal cycle cover of the overlap graph
and concatenates the strings obtained by breaking
each cycle at its minimum-overlap edge.
This algorithm was recently shown to achieve
a tight $3$-approximation.
Its reverse-complement extension, \textsf{MGREEDY-RC}, also
constructs an optimal cycle cover of the overlap graph
augmented with reverse complements, but its best known
approximation guarantee was $3.75$ (CPM 2026).
We extend this cycle-cover inequality to SCS-RC
and show that \textsf{MGREEDY-RC} achieves a
$3$-approximation, which is tight even over
the DNA alphabet.
The $3$-approximation guarantee for \textsf{MGREEDY-RC}
yields a $2.5$-approximation for \textsf{TGREEDY-RC},
which greedily merges the strings derived from the cycle cover.
By replacing this greedy merging step with an algorithm
based on a reduction to a maximum-weight Hamiltonian
path problem constrained to contain a given perfect matching,
we obtain a $(22/9+\varepsilon)$-approximation
for every fixed $\varepsilon>0$.
This improves upon the previous best ratio of $8/3$
(WABI 2026), establishing the best known polynomial-time
approximation guarantee for SCS-RC.
\end{abstract}

\section{Introduction}

The Shortest Common Superstring (SCS) problem
is a fundamental problem in string algorithms
and combinatorial optimization.
Given a set of strings, the goal is to find
a shortest string containing every input string
as a substring.
The problem has applications in various fields
\cite{Gevezes2014}.
One prominent application is genome assembly
\cite{li1990towards,KececiogluMyers1995}.
DNA sequencers produce short sequence reads,
from which the original genome must be reconstructed.
DNA sequences can be represented as strings over
the four-letter alphabet
$\{\texttt{A},\texttt{C},\texttt{G},\texttt{T}\}$,
corresponding to the four nucleotide bases
adenine, cytosine, guanine, and thymine.
Thus, under idealized assumptions, genome assembly
can be modeled as the SCS problem,
with the reads as input strings.
However, this SCS model does not capture the
fact that DNA consists of two complementary
strands and that the strand of origin of each
read is generally unknown
\cite{medvedev2009maximum,MyersJr+2016+126+132}.
Consequently, a reconstructed strand must
contain, for each read, either the read itself
or its reverse complement.
This motivates the Shortest Common Superstring
with Reverse Complements (SCS-RC) problem,
which asks for a shortest string containing,
for each input string, either that string
or its reverse complement.
Related problems incorporating reverse complements
have also been studied
\cite{conway2011succinct,CazauxCanovasRivals2016}.

Approximation algorithms for SCS have been
extensively studied
\cite{CPM1996.2+2/3approx,
Breslaure.1997.OverlapRotationLemma,
SIAM.2+1/2approx,SODA13.2+11/23approx}.
\textsf{MGREEDY} constructs an optimal cycle cover by selecting
maximum-overlap edges, then concatenates
strings obtained by breaking its cycles.
Its analysis relies on progressively stronger
upper bounds on the total overlap of
cycle-closing edges
\cite{blum1994linear,KAPLAN200513.greedy3.5n,
STOC22.ImprovedApproximateGuarantees,
englert_et_al:LIPIcs.ISAAC.2023.29}.
Recently, a tight bound was established
in \cite{chukhin2026tight},
yielding a tight $3$-approximation for
\textsf{MGREEDY}.
Combined with approximation algorithms for
the maximum asymmetric traveling salesman
problem
\cite{Kaplan.MAXATSP.2/3approx,
paluch_et_al:LIPIcs.STACS.2012.501},
this bound also yields a $7/3$-approximation
for SCS.
OpenAI recently released a preprint claiming
a polynomial-time $2$-approximation for SCS
using a non-greedy approach
\cite{OAI:A-Polynomial-Time-2-Approximation-for-Shortest-Common-Superstring-September-24-2026}.

In contrast, SCS-RC has received less attention
than SCS, despite better capturing the
double-stranded nature of DNA.
The work of \cite{JIANG1992195} introduced
SCS with reversals, noting that its results
apply to reverse complements.
It proved a $4$-approximation for the
reverse-complement extension of \textsf{MGREEDY},
which we call \textsf{MGREEDY-RC}
(\cref{alg:MGREEDY-RC}),
and mentioned a $3$-approximation
for \textsf{TGREEDY-RC} (\cref{alg:TGREEDY-RC}).
The work of \cite{yamano_et_al:LIPIcs.CPM.2026.15}
showed that a $(2+\alpha)$-approximation for
\textsf{MGREEDY-RC} yields a
$(2+\alpha/2)$-approximation for
\textsf{TGREEDY-RC}
(see \cref{lm:MGREEDY-RC-compress-framework}),
improving their ratios to $3.75$ and $2.875$,
respectively.
An $8/3$-approximation for SCS-RC was obtained
in \cite{yamano_et_al:LIPIcs.WABI.2026.2}
by carefully selecting where to break
the cycle-cover cycles.

\subsection*{Our Contributions}

We extend the tight cycle-cover inequality
of \cite{chukhin2026tight} to the
reverse-complement setting.
A key ingredient of their proof is a
Rauzy-graph argument that bounds the total
excess of cycle-closing overlap lengths
over the corresponding cycle periods.
We extend this argument by identifying
each substring with its reverse complement
in the Rauzy graph (see \cref{sec:non-self-rc}).
However, this alone does not handle cycles
whose periodic strings are cyclic rotations
of their reverse complements.
We therefore treat these cycles separately,
developing a new counting argument for
substrings that remain unchanged under
reverse complementation
(see \cref{sec:self-rc}).
These substring counts appear with opposite
signs in the two bounds and cancel when combined,
yielding the desired cycle-cover inequality.
We also establish a matching lower bound
even over the DNA alphabet by extending
the construction of \cite{chukhin2026tight}.

\begin{theorem} \label{thm:MGREEDY-RC-3-approx}
\textsf{MGREEDY-RC} is a $3$-approximation algorithm
for SCS-RC, and this bound is tight even over the DNA alphabet.
\end{theorem}
Combining \cref{thm:MGREEDY-RC-3-approx} with \cref{lm:MGREEDY-RC-compress-framework} yields \cref{thm:TGREEDY-RC-2.5-approx}.

\begin{theorem} \label{thm:TGREEDY-RC-2.5-approx}
\textsf{TGREEDY-RC} is a $2.5$-approximation algorithm
for SCS-RC.
\end{theorem}

We next consider maximum compression in SCS-RC,
which aims to maximize the difference between
the sum of the input string lengths and
the length of the resulting superstring.
A greedy $1/2$-approximation is known for
this problem \cite{FICI2016245}.
To improve this guarantee, we give an
approximation-preserving reduction to scaffolding,
which asks for a maximum-weight Hamiltonian path
containing a given perfect matching in a weighted graph.

\begin{theorem} \label{thm:scs-rc-max-compress-to-scaffolding}
A polynomial-time $\alpha$-approximation algorithm
for the scaffolding problem yields a polynomial-time
$\alpha$-approximation algorithm for maximum compression
in SCS-RC.
\end{theorem}
Applying \cref{thm:scs-rc-max-compress-to-scaffolding} to the
scaffolding algorithm of \cite{CHEN2018131} gives a compression
ratio of $5/9-\varepsilon$ for every fixed $\varepsilon>0$.
Combining this guarantee with
\cref{thm:MGREEDY-RC-3-approx,lm:MGREEDY-RC-compress-framework}
yields the following improvement over
the previous approximation ratio of $8/3$.

\begin{theorem} \label{thm:scs-rc-2.45-approx}
For every fixed $\varepsilon>0$,
SCS-RC admits a polynomial-time
$(22/9+\varepsilon)$-approximation algorithm.
\end{theorem}

\section{Preliminaries}

Throughout the paper, we consider strings over a finite
alphabet $\Gamma$ equipped with a \emph{complement mapping}
$\mathrm{cm}:\Gamma\to\Gamma$ that is an involution,
i.e., $\mathrm{cm}(\mathrm{cm}(x))=x$ for every $x\in\Gamma$.
An important example is the DNA alphabet
$\Gamma_{\mathrm{DNA}}=\{\texttt{A},\texttt{C},\texttt{G},\texttt{T}\}$,
where $\mathrm{cm}(\texttt{A})=\texttt{T}$ and
$\mathrm{cm}(\texttt{C})=\texttt{G}$ represent
the complementary base pairs.
For a string $s=a_1\ldots a_m\in\Gamma^m$,
its \emph{reverse complement}, denoted by $\mathrm{rc}(s)$,
is obtained by reversing $s$ and applying
$\mathrm{cm}$ to each character, that is,
$\mathrm{rc}(s)=\mathrm{cm}(a_m)\ldots\mathrm{cm}(a_1)$.
This general setting captures both standard reverse
complements of DNA strings and ordinary reversals
when the complement mapping is the identity.

For a set of strings $T$, let
$\mathrm{rc}(T)=\{\mathrm{rc}(t)\mid t\in T\}$.
A set of strings $T$ is \emph{RC-substring-free}
if, for any distinct $x,y\in T$, $x$ is not
a substring of either $y$ or $\mathrm{rc}(y)$.
By preprocessing, we may assume that the input set
$S$ is RC-substring-free.
We now define the
\emph{Shortest Common Superstring with Reverse Complements (SCS-RC)}
problem.

\begin{definition}[Shortest Common Superstring with Reverse Complements (SCS-RC)]
Given a set of strings $S=\{s_1,\ldots,s_n\}$
over $\Gamma$,
the objective is to find a shortest string
containing either $s$ or $\mathrm{rc}(s)$
as a substring for every $s\in S$.
\end{definition}

For each $s_i\in S$, we write
$s'_i\in\{s_i,\mathrm{rc}(s_i)\}$
to indicate a choice of orientation.
Let $\mathrm{ALG}(S)$ denote the superstring returned
by an approximation algorithm for SCS-RC, and let
$\mathrm{OPT}(S)$ denote an optimal superstring.
Unless stated otherwise, an $\alpha$-approximation
for SCS-RC satisfies
$|\mathrm{ALG}(S)|\le\alpha|\mathrm{OPT}(S)|$.
For maximum compression, we define
$\|S\|=\sum_{s\in S}|s|$ and say that an algorithm
achieves a $\beta$-approximation if
$\|S\|-|\mathrm{ALG}(S)|
\ge \beta(\|S\|-|\mathrm{OPT}(S)|)$.
In \cref{sec:max-compression-scaffolding} we give an approximation-preserving reduction
from maximum compression in SCS-RC to the
scaffolding problem defined below.

\begin{definition}[Scaffolding problem] \label{def:scaffolding-problem}
Given a complete undirected graph
$G=(V,E)$ with nonnegative edge weights and a perfect
matching $M\subseteq E$ of weight zero,
the objective is to find a maximum-weight Hamiltonian
path in $G$ containing $M$.
\end{definition}

We use the definitions of \emph{overlap} and \emph{prefix} from \cite{blum1994linear,JIANG1992195}.
For two not necessarily distinct strings $x$ and $y$,
their \emph{overlap}, denoted by $\mathrm{ov}(x,y)$,
is the longest string $v$ such that $x=uv$ and $y=vw$
for some nonempty strings $u$ and $w$.
We call $u$ the \emph{prefix} of $x$ with respect to $y$
and denote it by $\mathrm{pref}(x,y)$.
For strings $x_1,\ldots,x_r$,
we define their \emph{overlap merge} as
\[
\begin{aligned}
\langle x_1,\ldots,x_r\rangle
&=\mathrm{pref}(x_1,x_2)\cdots
  \mathrm{pref}(x_{r-1},x_r)x_r\\
&=\mathrm{pref}(x_1,x_2)\cdots
  \mathrm{pref}(x_{r-1},x_r)\mathrm{pref}(x_r,x_1)\mathrm{ov}(x_r,x_1).
\end{aligned}
\]

\subsection{Prefix and Overlap Graphs} \label{sec:prefix-and-overlap-graphs}

The \emph{prefix graph} of $S$, denoted by
$G_{\mathrm{pref}}(S)$, is a weighted complete directed
graph with vertex set $V=S\cup\mathrm{rc}(S)$ and
edge set $E=V\times V$, where each edge $(x,y)\in E$
has weight $|\mathrm{pref}(x,y)|$.
The \emph{overlap graph} $G_{\mathrm{ov}}(S)$ is defined
similarly, with edge weights $|\mathrm{ov}(x,y)|$
instead of $|\mathrm{pref}(x,y)|$.
The weight of a cycle is the sum of its edge weights,
with a self-loop having the weight of its single edge.

As observed in \cite{blum1994linear,JIANG1992195},
there exists an optimal solution to SCS-RC of the form
$\langle s'_{q_1},\ldots,s'_{q_n}\rangle$,
where $(q_1,\ldots,q_n)$ is a permutation of
$\{1,\ldots,n\}$.
Closing the corresponding path into a cycle yields
a cycle of weight $|\mathrm{OPT}(S)|
-|\mathrm{ov}(s'_{q_n},s'_{q_1})|
\le |\mathrm{OPT}(S)|$.
Thus, the minimum weight of a cycle that visits exactly
one vertex from $\{s,\mathrm{rc}(s)\}$ for each $s\in S$
is a lower bound on $|\mathrm{OPT}(S)|$.
We relax this problem to finding a minimum-weight
\emph{cycle cover}.

A \emph{cycle cover} of the prefix or overlap graph
consists of vertex-disjoint cycles, possibly
including self-loops, that visit exactly
one vertex from $\{s,\mathrm{rc}(s)\}$ for each $s\in S$.
Following \cite{JIANG1992195}, this definition extends
the standard notion to account for reverse complements.
Since $|\mathrm{pref}(x,y)|=|x|-|\mathrm{ov}(x,y)|$,
the prefix and overlap weights of any cycle cover
sum to $\|S\|$.
Thus, a minimum-weight cycle cover of $G_{\mathrm{pref}}(S)$
is a maximum-weight cycle cover of $G_{\mathrm{ov}}(S)$,
and vice versa.
Let $W$ denote the minimum weight of a cycle cover
of $G_{\mathrm{pref}}(S)$.
Then
\begin{equation}
\label{eq:cycle-cover-lower-bound}
W \le |\mathrm{OPT}(S)|.
\end{equation}

\subsection{Greedy Algorithms}

\begin{algorithm}[t]
\DontPrintSemicolon
\caption{\textsf{MGREEDY-RC}}
\label{alg:MGREEDY-RC}
$T \leftarrow \emptyset$\;
\While{$S \neq \emptyset$}{
    $P \leftarrow \{(x,y)\in (S \cup \mathrm{rc}(S)) \times (S \cup \mathrm{rc}(S)) \mid
    x=y \text{ or } y\notin\{x,\mathrm{rc}(x)\}\}$\;
    Choose $(x,y)\in P$ maximizing $|\mathrm{ov}(x,y)|$\;
    \eIf{$x=y$}{
        $S \leftarrow S\setminus\{x,\mathrm{rc}(x)\}$, \quad 
        $T \leftarrow T\cup\{x\}$\;
    }{
        $S \leftarrow
        (S\setminus\{x,\mathrm{rc}(x),y,\mathrm{rc}(y)\})
        \cup\{\langle x,y\rangle\}$\;
    }
}
\Return{the concatenation of the strings in $T$}\;
\end{algorithm}

\begin{algorithm}[t]
\DontPrintSemicolon
\caption{\textsf{GREEDY-RC}}
\label{alg:GREEDY-RC}
\While{$|S|>1$}{
    $P \leftarrow \{(x,y)\in (S \cup \mathrm{rc}(S)) \times (S \cup \mathrm{rc}(S)) \mid
    y\notin\{x,\mathrm{rc}(x)\}\}$\;
    Choose $(x,y)\in P$ maximizing $|\mathrm{ov}(x,y)|$\;
    $S \leftarrow
    (S\setminus\{x,\mathrm{rc}(x),y,\mathrm{rc}(y)\})
    \cup\{\langle x,y\rangle\}$\;
}
\Return{the unique element of $S$}\;
\end{algorithm}

\begin{algorithm}[t]
\DontPrintSemicolon
\caption{\textsf{TGREEDY-RC}}
\label{alg:TGREEDY-RC}
Run \textsf{MGREEDY-RC} without its final
concatenation step, obtaining $T$\;
\Return{$\textsf{GREEDY-RC}(T)$}\;
\end{algorithm}

The \textsf{MGREEDY-RC} algorithm is shown in
\cref{alg:MGREEDY-RC}.
The following lemma shows that greedy merges
preserve the RC-substring-free property
established by preprocessing.
\begin{lemma}
\label{lm:substring-free-invariant}
Throughout \textsf{MGREEDY-RC}, the
working set $S$ remains RC-substring-free.
\end{lemma}
\begin{proof}
Initially, $S$ is RC-substring-free by preprocessing,
and removing a string preserves this property.
Suppose the first violation occurs when
$x$ and $y$ are merged with overlap
length $k$, producing $t=\langle x,y\rangle$
that contains $z\in\{s,\mathrm{rc}(s)\}$
for some remaining $s\in S$.
Since $\{x,y,s\}$ was RC-substring-free,
the occurrence of $z$ starts strictly inside
$x$ and extends beyond its end.
This yields an overlap between $x$ and $z$
larger than $k$, contradicting the greedy choice.
\end{proof}
As observed in \cite{JIANG1992195}, \textsf{MGREEDY-RC} can be viewed
as constructing a cycle cover.
Each string $\langle s'_{i_1},\ldots,s'_{i_r}\rangle$
corresponds to a path $(s'_{i_1},\ldots,s'_{i_r})$
in the prefix or overlap graph.
Merging two strings corresponds to joining their paths
by an edge from the last vertex of the first path
to the first vertex of the second.
Taking the reverse complement of a string corresponds
to reversing its path and taking the reverse complement
of each vertex.
Finally, moving a string from $S$ to $T$ corresponds
to closing its path into a cycle.
Thus, the strings in $T$ correspond to the cycles
of the resulting cycle cover.
Fix an arbitrary execution of \textsf{MGREEDY-RC} on $S$,
and let $\mathcal{C}(S)$ denote the cycle cover it produces.

\begin{lemma}[\cite{JIANG1992195}]
$\mathcal{C}(S)$ is a minimum-weight cycle cover
of $G_{\mathrm{pref}}(S)$.
\end{lemma}
For each $C\in\mathcal{C}(S)$, write
$C=(s'_{i_1},\ldots,s'_{i_r},s'_{i_1})$
so that $(s'_{i_r},s'_{i_1})$ is its closing edge.
We define the \emph{representative string}
$\sigma(C)$ and the \emph{cycle string} $p(C)$ as
\[
\sigma(C)
=\langle s'_{i_1},\ldots,s'_{i_r}\rangle,\quad
p(C)
=\mathrm{pref}(s'_{i_1},s'_{i_2})\cdots
\mathrm{pref}(s'_{i_{r-1}},s'_{i_r})
\mathrm{pref}(s'_{i_r},s'_{i_1}),
\]
respectively.
Since \textsf{MGREEDY-RC} selects overlaps in
nonincreasing order of weight, the closing edge
of $C$ has minimum overlap weight among its edges.
Let $w(C)$ denote the total prefix weight of $C$,
and let $o(C)$ denote the overlap weight of its
closing edge.
Then, $|p(C)|=w(C)$ and
$|\sigma(C)|=w(C)+o(C)$.
Recall that $W=\sum_{C\in\mathcal{C}(S)}w(C)$,
and let $O=\sum_{C\in\mathcal{C}(S)}o(C)$.
Since \textsf{MGREEDY-RC} concatenates the
representative strings, the length of its output is
\begin{equation} \label{eq:MGREEDY-RC-output}
\sum_{C\in\mathcal{C}(S)}|\sigma(C)|=W+O.
\end{equation}
The \textsf{GREEDY-RC} algorithm shown in \cref{alg:GREEDY-RC} serves as the compression step in the \textsf{TGREEDY-RC} algorithm shown in \cref{alg:TGREEDY-RC}.

\begin{lemma}[\cite{FICI2016245}]
\textsf{GREEDY-RC} achieves an approximation ratio
of $1/2$ for maximum compression in SCS-RC.
This ratio is tight.
\end{lemma}

\begin{lemma}[\cite{yamano_et_al:LIPIcs.CPM.2026.15}]
\label{lm:MGREEDY-RC-compress-framework}
If \textsf{MGREEDY-RC} achieves an approximation ratio
of $2+\alpha$, combining it with an algorithm
that achieves a compression ratio of $\beta$
for SCS-RC yields a $(2+(1-\beta)\alpha)$-approximation
algorithm.
Since \textsf{GREEDY-RC} achieves a compression ratio
of $1/2$, \textsf{TGREEDY-RC} achieves an approximation
ratio of $2+\alpha/2$.
\end{lemma}
Thus, our 3-approximation guarantee for
\textsf{MGREEDY-RC} immediately implies that
\textsf{TGREEDY-RC} achieves a 2.5-approximation.
Furthermore, replacing the \textsf{GREEDY-RC}
compression step with the scaffolding-based
$(5/9-\varepsilon)$-approximation algorithm for
maximum compression improves the approximation
ratio to $22/9+\varepsilon$.

\subsection{Periodicity and Lyndon Words}

For nonempty strings $x,y$, we call $y$ a
\emph{periodic root} of $x$ if $x=y^ky'$ for some
integer $k\ge 1$ and a possibly empty proper prefix
$y'$ of $y$. Then $|y|$ is a \emph{period} of $x$.
Two strings $u$ and $v$ are \emph{equivalent}
if one is a cyclic rotation of the other,
and \emph{inequivalent} otherwise.
We write $u\sim v$ and $u\not\sim v$, respectively.
A nonempty string $x$ is \emph{primitive} if
$x\ne y^k$ for all nonempty strings $y$ and
integers $k\ge 2$.
We use the following known properties.

\begin{lemma}[\cite{blum1994linear}]
\label{lm:cycle-strings-periodic-root}
For every cycle $C\in\mathcal{C}(S)$,
$p(C)$ is the shortest periodic root of $\sigma(C)$
and hence is primitive.
\end{lemma}

\begin{lemma}[\cite{JIANG1992195,yamano_et_al:LIPIcs.CPM.2026.15}]
\label{lm:cycle-strings-non-equiv}
For any two distinct cycles $C,D\in\mathcal{C}(S)$,
we have $p(C)\not\sim p(D)$ and
$p(C)\not\sim\mathrm{rc}(p(D))$.
\end{lemma}
A \emph{Lyndon word} \cite{lyndon1954burnside}
is a nonempty string that is lexicographically
smaller than all its nontrivial cyclic rotations.
Every Lyndon word $z$ has minimum period $|z|$.
Indeed, if $z$ had a shorter period $q$,
we could write $z=uv=wu$ with
$|v|=|w|=q$.
Comparing $z$ with its cyclic rotations
$uw$ and $vu$ would imply both
$v<w$ and $w<v$, a contradiction.

For each cycle $C\in\mathcal{C}(S)$, let $z(C)$
denote the lexicographically smallest among all
cyclic rotations of $p(C)$ and $\mathrm{rc}(p(C))$.
Since both $p(C)$ and $\mathrm{rc}(p(C))$ are
primitive by \cref{lm:cycle-strings-periodic-root},
$z(C)$ is a Lyndon word.

\section{Proving that \textsf{MGREEDY-RC} is 3-Approximate} \label{sec:MGREEDY-RC-3-upperbound}

By \eqref{eq:cycle-cover-lower-bound} and
\eqref{eq:MGREEDY-RC-output},
\cref{lm:O<=W+OPT} implies that
\textsf{MGREEDY-RC} is a $3$-approximation algorithm.

\begin{lemma}
\label{lm:O<=W+OPT}
$O\le W+|\mathrm{OPT}(S)|$.
\end{lemma}
The remainder of this section is devoted to proving
\cref{lm:O<=W+OPT}.
Since cycles $C\in\mathcal{C}(S)$ with $o(C)\le w(C)$
contribute nonpositively to $O-W$, we focus on
the set $\mathcal{H}(S)=
\{C\in\mathcal{C}(S)\mid o(C)>w(C)\}$.
To account for reverse complements, we further
partition $\mathcal{H}(S)$ into
\[
\mathcal{H}_{\mathrm{self}}(S)
=\{C\in\mathcal{H}(S)\mid
p(C)\sim\mathrm{rc}(p(C))\},\quad
\mathcal{H}_{\mathrm{nonself}}(S)
=\{C\in\mathcal{H}(S)\mid
p(C)\not\sim\mathrm{rc}(p(C))\}.
\]
For a nonempty string $u$ and an integer $m\ge 0$,
let $\mathrm{Sub}_m(u)$ and $\mathrm{CSub}_m(u)$
denote the sets of all length-$m$ substrings
of $u$ and $u^\infty$, respectively.
For $m=0$, both sets consist only of the empty string
$\varepsilon$.
To account for reverse complements, we further define
\[
\mathrm{Sub}_m^R(u)
=\bigl\{\{w,\mathrm{rc}(w)\}
\mid w\in\mathrm{Sub}_m(u)\bigr\},\quad
\mathrm{SelfSub}_m^R(u)
=\{w\in\mathrm{Sub}_m(u)
\mid w=\mathrm{rc}(w)\}.
\]
To prove \cref{lm:O<=W+OPT}, we establish
the following two lemmas.

\begin{restatable}{lemma}{LmNonSelfCycleSum}
\label{lm:nonself-cycle-sum}
The cycles in $\mathcal{H}_{\mathrm{nonself}}(S)$
satisfy
\[
\sum_{C\in\mathcal{H}_{\mathrm{nonself}}(S)}
(o(C)-w(C)+2)
\le |\mathrm{OPT}(S)|-\sum_{m=1}^{|\mathrm{OPT}(S)|}|\mathrm{SelfSub}_m^R(\mathrm{OPT}(S))|.
\]
\end{restatable}

\begin{restatable}{lemma}{LmSelfCycleSum}
\label{lm:self-cycle-sum}
The cycles in $\mathcal{H}_{\mathrm{self}}(S)$
satisfy
\[
\sum_{C\in\mathcal{H}_{\mathrm{self}}(S)}
(o(C)-w(C)+2)
\le\sum_{m=1}^{|\mathrm{OPT}(S)|}|\mathrm{SelfSub}_m^R(\mathrm{OPT}(S))|.
\]
\end{restatable}
Summing the inequalities in
\cref{lm:nonself-cycle-sum,lm:self-cycle-sum}
gives
$\sum_{C\in\mathcal{H}(S)}
(o(C)-w(C)+2)\le|\mathrm{OPT}(S)|$.
Since $o(C)-w(C)\le 0$ for every
$C\in (\mathcal{C}(S) \setminus \mathcal{H}(S))$, we obtain
\[
O-W
=\sum_{C\in\mathcal{C}(S)}(o(C)-w(C))
\le\sum_{C\in\mathcal{H}(S)}
(o(C)-w(C)+2)
\le|\mathrm{OPT}(S)|.
\]
Thus, \cref{lm:O<=W+OPT} follows from
\cref{lm:nonself-cycle-sum,lm:self-cycle-sum},
so it remains to prove these two lemmas.

\subsection{Proof of Lemma \ref{lm:nonself-cycle-sum}} \label{sec:non-self-rc}

\begin{figure}[t]
    \centering
    \includegraphics[width=0.9\linewidth]{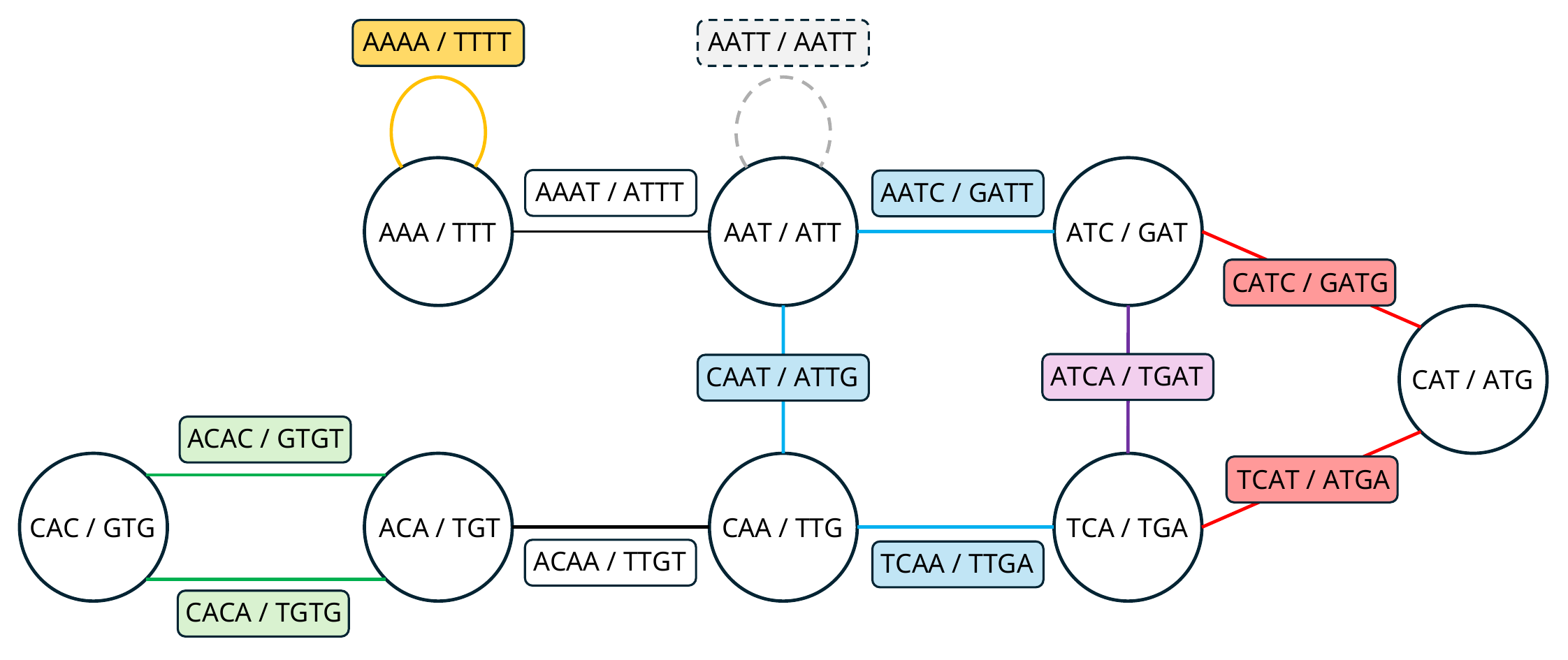}
\caption{
An example of the graph constructed in
\cref{lm:rauzy-graphs}, where
$u=\texttt{AAAATCATCAATTGTGT}$, $m=4$, and
$Z=\{z_1,z_2,z_3,z_4\}
=\{\texttt{A},\texttt{AC},\texttt{ATG},\texttt{AATC}\}$.
Each $[w]_R$ is displayed as $w/\mathrm{rc}(w)$.
Here, $y_i=z_i$ for $i\ne 3$, while
$y_3=\texttt{ATC}$ is the smallest cyclic
rotation of $\mathrm{rc}(z_3)=\texttt{CAT}$.
We have $|\mathrm{Sub}_{m}^R(u)|=12$,
$|\mathrm{Sub}_{m-1}^R(u)|=8$, and
$\mathrm{SelfSub}_{m}^R(u)=\{\texttt{AATT}\}$.
The gray dashed self-loop corresponds to
$\texttt{AATT}=\mathrm{rc}(\texttt{AATT})$
and is removed when constructing $G_m$.
The closed trails of $y_1,\ldots,y_4$
are shown in yellow, green, red, and blue,
respectively.
The purple edge $[\texttt{ATCA}]_R$
is shared by the last two trails.
Removing the edges
$[\texttt{AAAA}]_R$, $[\texttt{AATC}]_R$,
$[\texttt{ACAC}]_R$, and $[\texttt{ATCA}]_R$
leaves $G_m$ connected.
}
    \label{fig:rauzy-graph}
\end{figure}

We first prove \cref{lm:rauzy-graphs}, adapting the
undirected Rauzy-graph argument of
\cite[Lemma 3.2]{chukhin2026tight} to the
reverse-complement setting.
The adaptation identifies substrings with their reverse complements, treating those fixed by reverse complementation separately.

\begin{lemma} \label{lm:rauzy-graphs}
Let $u$ be a string and $1\le m\le |u|$ be an integer.
Let $Z$ be a set of Lyndon words such that
$z\not\sim\mathrm{rc}(z')$ for all $z,z'\in Z$.
Suppose that, for every $z\in Z$, $|z|\le m$ and
$\mathrm{CSub}_m(z)
\subseteq
\mathrm{Sub}_m(u)\cup\mathrm{Sub}_m(\mathrm{rc}(u))$.
Then,
\[
|Z|\le
|\mathrm{Sub}_m^R(u)|
-|\mathrm{Sub}_{m-1}^R(u)|
-|\mathrm{SelfSub}_m^R(u)|+1.
\]
\end{lemma}
\begin{proof}
For a string $w$, let $[w]_R=\{w,\mathrm{rc}(w)\}$.
Form an undirected multigraph with vertex set
$\mathrm{Sub}_{m-1}^R(u)$.
For each $[w]_R\in\mathrm{Sub}_m^R(u)$,
add an edge joining $[x]_R$ and $[y]_R$,
where $x$ and $y$ are the length-$(m-1)$
prefix and suffix of $w$, respectively.
This is well-defined since replacing $w$ with
$\mathrm{rc}(w)$ swaps the endpoints.
Consecutive length-$m$ substrings of $u$ trace
a walk visiting every vertex, so the graph is connected.
If $w=\mathrm{rc}(w)$, the corresponding edge
is a self-loop.
Removing such self-loops preserves connectivity.
Let $G_m$ denote the resulting graph.
Thus,
\[
|V(G_m)|=|\mathrm{Sub}_{m-1}^R(u)|,\quad
|E(G_m)|=|\mathrm{Sub}_m^R(u)|
-|\mathrm{SelfSub}_m^R(u)|.
\]
We show that removing $|Z|$ edges from $G_m$
preserves connectivity.
We order the edges of $G_m$ lexicographically,
using $\min\{w,\mathrm{rc}(w)\}$ as the label
of each edge corresponding to $[w]_R$.
By assumption, for each $z_i\in Z$, the graph
$G_m$ contains all edges corresponding to
$[w]_R$ with $w\in\mathrm{CSub}_m(z_i)$.
Let $y_i$ be the lexicographically smallest
cyclic rotation of either $z_i$ or
$\mathrm{rc}(z_i)$.
Since $m\ge |z_i|$, $z_i$ is primitive, and
$z_i\not\sim\mathrm{rc}(z_i)$,
the $|z_i|$ cyclic length-$m$ windows of
$y_i^\infty$ correspond to distinct edges
of $G_m$, forming a closed trail in cyclic order.
Let $b_i$ be the lexicographically smallest
edge of this trail.

For distinct $z_i,z_j\in Z$, suppose that $b_i=b_j$.
The common edge label is the length-$m$ prefix
of both $y_i^\infty$ and $y_j^\infty$.
Since Lyndon words have no shorter periods,
we obtain $|y_i|=|y_j|$, and hence $y_i=y_j$.
This implies $z_i\sim z_j$ or
$z_i\sim\mathrm{rc}(z_j)$, a contradiction.
Thus, the edges $b_1,\ldots,b_{|Z|}$ are
pairwise distinct.
Reindex the pairs $(z_i,b_i)$ so that
$b_1<\cdots<b_{|Z|}$.
Removing $b_1,\ldots,b_{|Z|}$ in this order
preserves connectivity: each removed edge is
either a self-loop or has a bypass consisting
of larger, undeleted edges of its closed trail.
\Cref{fig:rauzy-graph} illustrates the construction
of $G_m$ and the selection of edges to be removed
for a given $Z$.
Since removing $|Z|$ edges from $G_m$ preserves
connectivity, we have
$|E(G_m)|-|Z|\ge |V(G_m)|-1$.
Therefore,
\[
|Z|
\le |E(G_m)|-|V(G_m)|+1
=|\mathrm{Sub}_m^R(u)|
-|\mathrm{Sub}_{m-1}^R(u)|
-|\mathrm{SelfSub}_m^R(u)|+1.
\qedhere
\]
\end{proof}
To apply \cref{lm:rauzy-graphs} to the strings $z(C)$ for each cycle $C \in \mathcal{H}_{\mathrm{nonself}}(S)$, we prove \cref{lm:zc-window}.
\begin{lemma} \label{lm:zc-window}
For every cycle $C \in \mathcal{C}(S)$ and $1 \le m \le o(C)+1$,
\[
\mathrm{CSub}_{m}(z(C)) 
\subseteq\bigcup_{s\in S}
(\mathrm{Sub}_m(s)\cup\mathrm{Sub}_m(\mathrm{rc}(s)))
\subseteq \mathrm{Sub}_{m}(\mathrm{OPT}(S)) \cup \mathrm{Sub}_m(\mathrm{rc}(\mathrm{OPT}(S))).
\]
\end{lemma}
\begin{proof}
The second inclusion follows from the
definition of SCS-RC, so it suffices to prove
the first inclusion.
Let $C=(s'_{i_1},\ldots,s'_{i_r},s'_{i_{r+1}})$,
where $s'_{i_{r+1}}=s'_{i_1}$.
By the definition of the overlap merge,
each block $\mathrm{pref}(s'_{i_j},s'_{i_{j+1}})$
of $p(C)$ is followed in $\sigma(C)$
by at least $o(C)$ characters of $s'_{i_j}$.
Each of the first $w(C)$ positions of
$\sigma(C)$ lies in one of these blocks.
Thus, every length-$m$ substring starting there
is contained in the corresponding $s'_{i_j}$,
since $m\le o(C)+1$.
Since $|p(C)|=w(C)$, these positions
cover all of $\mathrm{CSub}_m(p(C))$.
The same holds for
$\mathrm{CSub}_m(\mathrm{rc}(p(C)))$
by taking reverse complements.
Since $z(C)$ is a cyclic rotation of either
$p(C)$ or $\mathrm{rc}(p(C))$,
the first inclusion follows.
\end{proof}
Now we are ready to prove \cref{lm:nonself-cycle-sum}, which we restate here.
\LmNonSelfCycleSum*

\begin{proof}
Let $u=\mathrm{OPT}(S)$.
By \cref{lm:cycle-strings-non-equiv}, the words
$z(C)$ for $C\in\mathcal{H}_{\mathrm{nonself}}(S)$
are pairwise distinct and satisfy
$z(C)\not\sim\mathrm{rc}(z(D))$
for all $C,D\in\mathcal{H}_{\mathrm{nonself}}(S)$.
For each $1\le m\le |u|$, let
$Z_m=\{z(C)\mid
C\in\mathcal{H}_{\mathrm{nonself}}(S),\;
w(C)\le m\le o(C)+1\}$.
Since $|z(C)|=w(C)$, \cref{lm:zc-window}
ensures that $Z_m$ satisfies the assumptions
of \cref{lm:rauzy-graphs}.
Thus,
\[
|Z_m|\le
|\mathrm{Sub}_m^R(u)|
-|\mathrm{Sub}_{m-1}^R(u)|
-|\mathrm{SelfSub}_m^R(u)|+1.
\]
Since overlaps are proper, $o(C)+1\le |u|$.
In the sum $\sum_{m=1}^{|u|}|Z_m|$,
each $z(C)$ is counted exactly
$o(C)-w(C)+2$ times.
Therefore, we obtain
\begin{align*}
&\sum_{C\in\mathcal{H}_{\mathrm{nonself}}(S)}
(o(C)-w(C)+2)
=\sum_{m=1}^{|u|}|Z_m| \\
&\le
\sum_{m=1}^{|u|}
\bigl(
|\mathrm{Sub}_m^R(u)|
-|\mathrm{Sub}_{m-1}^R(u)|
-|\mathrm{SelfSub}_m^R(u)|+1
\bigr)\\
&=|\mathrm{Sub}_{|u|}^R(u)|
-|\mathrm{Sub}_0^R(u)|
+|u|-\sum_{m=1}^{|u|}|\mathrm{SelfSub}_m^R(u)|
=|u|-\sum_{m=1}^{|u|}|\mathrm{SelfSub}_m^R(u)|,
\end{align*}
where the last equality follows from
$|\mathrm{Sub}_{|u|}^R(u)|
=|\mathrm{Sub}_0^R(u)|=1$.
\end{proof}

\subsection{Proof of Lemma \ref{lm:self-cycle-sum}} \label{sec:self-rc}

For each $C\in\mathcal{H}_{\mathrm{self}}(S)$,
we have $p(C)\sim\mathrm{rc}(p(C))$,
meaning that the reverse complement of $p(C)$
is a cyclic rotation of $p(C)$.
Let $X=p(C)^\infty$, extended periodically
to all integer indices, and write $X[i]$
for its $i$-th character.
Then there exists an integer $0\le k<w(C)$
such that
\[
X[k+i]=\mathrm{cm}(X[w(C)+1-i])
\qquad (1\le i\le w(C)).
\]
Thus, $X$ is invariant under reflection about
the axis $(w(C)+k+1)/2$ followed by
complementation of each character.
By periodicity, the same symmetry holds about
every axis $(t \cdot w(C)+k+1)/2$
for $t\in\mathbb{Z}$.
We now consider the representative string
$\sigma(C)=X[1..w(C)+o(C)]$,
where $X[i..j]=X[i]\cdots X[j]$.
Let $\alpha$ and $\beta$ be two consecutive
reflection-complement symmetry axes bracketing
the center of $\sigma(C)$.
Then, $\beta=\alpha+w(C)/2$ and
\[
\alpha\le\frac{1+w(C)+o(C)}{2}\le\beta.
\]
We choose an occurrence of either $z(C)$ or
$\mathrm{rc}(z(C))$ on $X$ whose center lies
between $\alpha$ and $\beta$.
Since $p(C)\sim z(C)$,
$z(C)$ occurs in $X$, with occurrences repeating
every $w(C)$ positions.
Let $c$ be the center of the first such occurrence
at or to the right of $\alpha$.
Then $c \in [\alpha, \alpha+w(C)]$.
If $c\le\beta$, we choose this occurrence.
Otherwise, reflecting it about $\beta$ and
complementing its characters yields an occurrence
of $\mathrm{rc}(z(C))$ centered at
$2\beta-c\in[\alpha,\beta]$.
Thus, the required occurrence exists.
Let $[Z_1,Z_2]$ denote the interval occupied
by this occurrence.

\begin{figure}[b]
\centering
\includegraphics[width=0.9\linewidth]{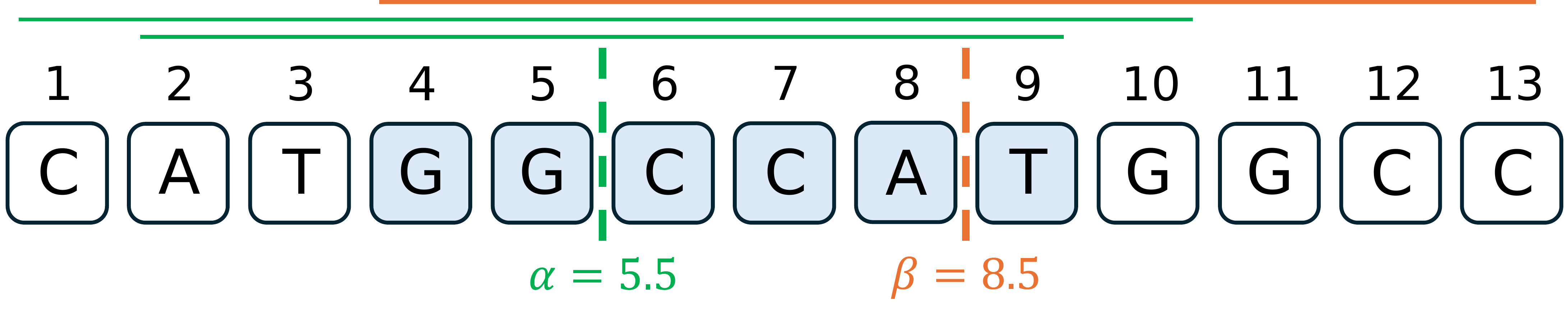}
\caption{
An example of selecting $o(C)-w(C)+2=3$
substrings centered at $\alpha=5.5$ or $\beta=8.5$.
Here, $p(C)=\texttt{CATGGC}$, $w(C)=6$, and $o(C)=7$.
The interval $[Z_1,Z_2]=[4,9]$, highlighted in blue,
corresponds to $\mathrm{rc}(z(C))=\texttt{GGCCAT}$.
The substrings centered at $\alpha$ occupy
$[1,10]$ and $[2,9]$ (green lines),
while the substring centered at $\beta$ occupies
$[4,13]$ (orange line).
}
\label{fig:self-rc-example}
\end{figure}

We consider substrings of $\sigma(C)$ whose
intervals contain $[Z_1,Z_2]$, lie within
$[1,w(C)+o(C)]$, and are centered at
either $\alpha$ or $\beta$.
An example of selecting such substrings is shown in \cref{fig:self-rc-example}.
By the reflection-complement symmetry of $X$,
each such substring equals its reverse complement.
We prove the following three properties:
\begin{enumerate}
\item There are at least $o(C)-w(C)+2$
such substrings for each
$C\in\mathcal{H}_{\mathrm{self}}(S)$.
\item All these substrings are pairwise distinct,
even across different cycles.
\item Each such substring occurs in an
original input string.
\end{enumerate}

These properties imply that there are at least
$\sum_{C\in\mathcal{H}_{\mathrm{self}}(S)}
(o(C)-w(C)+2)$ distinct elements in
$\bigcup_{m=1}^{|\mathrm{OPT}(S)|}
\mathrm{SelfSub}_m^R(\mathrm{OPT}(S))$,
which proves \cref{lm:self-cycle-sum}.
We prove them in order.

\subparagraph*{Property 1}
The intervals centered at $\alpha$ are $[i, 2\alpha-i]$ where $1\le i$ and $Z_2 \le 2\alpha-i$.
The intervals centered at $\beta$ are $[2\beta-j,j]$ where $2\beta-j \le Z_1$ and $j\le w(C)+o(C)$.
Thus, their numbers are
$\max\{0,2\alpha-Z_2\}$ and
$\max\{0,w(C)+o(C)-2\beta+Z_1+1\}$, respectively.
Since $\beta-\alpha=w(C)/2$ and
$Z_2-Z_1+1=w(C)$, their sum is at least
\[
(2\alpha-Z_2)+(w(C)+o(C)-2\beta+Z_1+1)
=o(C)-w(C)+2.
\]

\subparagraph*{Property 2}
We first show that the selected substrings from
the same cycle are pairwise distinct.
Those centered at the same axis have different lengths,
so a repetition is possible only between
equal-length substrings centered at $\alpha$ and $\beta$.
If $w(C)$ is odd, one axis is an integer
and the other a half-integer.
Hence, substrings centered at these axes have
opposite length parities and cannot be equal.
If $w(C)$ is even, suppose that two such
substrings are equal.
Their occurrences in $\sigma(C)$ are shifted by
$\beta-\alpha=w(C)/2$.
Since both contain $[Z_1,Z_2]$, their lengths
are at least $w(C)$.
Their equality then implies that $\sigma(C)$ has
period $w(C)/2$,
contradicting \cref{lm:cycle-strings-periodic-root}.
Thus, all selected substrings from
the same cycle are distinct.

It remains to show that substrings selected from
distinct cycles are also distinct.
Each selected substring from a cycle $C$ has
period $w(C)$, since it occurs in $p(C)^\infty$.
Moreover, it contains either $z(C)$ or
$\mathrm{rc}(z(C))$, both of which have
minimum period $w(C)$.
Thus, its minimum period is exactly $w(C)$.
Suppose that the same substring is selected
from two distinct cycles $C$ and $D$.
Then $w(C)=w(D)$.
Since the substring has length at least $w(C)$
and occurs in both $p(C)^\infty$ and $p(D)^\infty$,
we obtain $p(C)\sim p(D)$,
contradicting \cref{lm:cycle-strings-non-equiv}.

\subparagraph*{Property 3}
Suppose that a selected substring $v$, occupying
$[l,r]$ in $\sigma(C)$, does not occur in
any original input string or its reverse complement.
Consider the first merge in the execution of
\textsf{MGREEDY-RC} that creates this occurrence
of $v$ by merging two strings $x$ and $y$.
Let $x$ and $y$ occupy $[a_1,b_1]$ and
$[a_2,b_2]$, respectively, in $\sigma(C)$,
where $a_1<a_2$ and $b_1<b_2$.
Since neither interval contains $[l,r]$
but their merge does, we have
$a_1\le l<a_2$ and $b_1<r\le b_2$.
Let $\gamma=(l+r)/2\in\{\alpha,\beta\}$,
$c_x=(a_1+b_1)/2$, and $c_y=(a_2+b_2)/2$.
Then $c_x<\gamma<c_y$.
Reflecting $x$ about $\gamma$ and complementing
its characters yields an occurrence of
$\mathrm{rc}(x)$ on $X$ with center
$c'_x=2\gamma-c_x$.
Since $|c_y-c'_x|<c_y-c_x$ and
$\{x,y\}$ is RC-substring-free by
\cref{lm:substring-free-invariant},
this yields a larger overlap between
$\mathrm{rc}(x)$ and $y$ in some order,
contradicting the greedy choice of
\textsf{MGREEDY-RC}.

\section{Tight Lower Bound for \textsf{MGREEDY-RC}}

We show that the approximation ratio of $3$
established for \textsf{MGREEDY-RC} in
\cref{sec:MGREEDY-RC-3-upperbound} is tight,
by extending the lower-bound instance of
\cite[Section 3.2]{chukhin2026tight}
to the reverse-complement setting over
the DNA alphabet.

\begin{lemma}
For every fixed $\delta>0$, there exists an RC-substring-free
set $S$ of DNA strings such that every execution of
\textsf{MGREEDY-RC} returns a string of length
at least $(3-\delta)|\mathrm{OPT}(S)|$.
\end{lemma}
\begin{proof}

For integers $p\ge 2$ and $m \ge 2$, let
\[
s_p=\texttt{A}^{p-1}\texttt{C}\texttt{A}^{p-1}
\texttt{C}\texttt{A}^{p-2}
\qquad S=\{s_2,\ldots,s_m\}.
\]
The two occurrences of \texttt{C} in $s_p$ are
exactly $p$ positions apart.
Thus, $s_p$ cannot be a substring of $s_q$
for $p\ne q$, and the same holds for their
reverse complements.
Moreover, the strings in $S$ use only
$\{\texttt{A},\texttt{C}\}$, whereas those in
$\mathrm{rc}(S)$ use only $\{\texttt{T},\texttt{G}\}$.
Hence, $S$ is RC-substring-free, and all overlaps
between $S$ and $\mathrm{rc}(S)$ are empty,
so we only need to consider overlaps within $S$.

The string $\texttt{A}^{p-1}\texttt{C}\texttt{A}^{p-2}$
is a common proper prefix and suffix of $s_p$
of length $2p-2$.
Any longer self-overlap would require a shift
of fewer than $p$ positions, which cannot
align the occurrences of \texttt{C}.
Thus, $|\mathrm{ov}(s_p,s_p)|=2p-2$ and
$|\mathrm{pref}(s_p,s_p)|=p$.
For $p\ne q$, any suffix of $s_p$ longer than
$2p-2$ contains both occurrences of \texttt{C}
and thus cannot match a prefix of $s_q$.
An overlap of length exactly $2p-2$ is also
impossible, since the first \texttt{C} occurs
at position $p$ in this suffix of $s_p$,
but at position $q$ in $s_q$.
Hence, $|\mathrm{ov}(s_p,s_q)|\le 2p-3$
for $p\ne q$.
Therefore, \textsf{MGREEDY-RC} always selects
self-overlaps and produces a cycle cover
$\mathcal{C}(S)$ consisting entirely of self-loops.
Let $L_m$ denote the length of the superstring
returned by \textsf{MGREEDY-RC}.
Then,
\[
W=\sum_{p=2}^m|\mathrm{pref}(s_p,s_p)|
=\frac{m(m+1)}{2}-1,\quad
L_m=\sum_{p=2}^m|s_p|=3W-2(m-1).
\]
Since $|\mathrm{ov}(s_p,s_{p-1})|=2p-3$,
we have $|\mathrm{pref}(s_p,s_{p-1})|=p+1$.
Therefore,
\[
|\mathrm{OPT}(S)|
\le|\langle s_m,\ldots,s_2\rangle|
=|s_2|+\sum_{p=3}^m
|\mathrm{pref}(s_p,s_{p-1})|
=W+m.
\]
Thus, the approximation ratio satisfies
\[
\frac{L_m}{|\mathrm{OPT}(S)|} \ge \frac{3W-2(m-1)}{W+m} = 3 - \frac{5m-2}{W+m}.
\]
Since $W=\Theta(m^2)$, for every fixed $\delta>0$, the ratio is at least $3-\delta$
for sufficiently large $m$.
\end{proof}

Hence, the approximation ratio of $3$ is tight
even over the DNA alphabet, completing the proof
of \cref{thm:MGREEDY-RC-3-approx}.
Combining the 3-approximation guarantee with
\cref{lm:MGREEDY-RC-compress-framework}
yields a 2.5-approximation for \textsf{TGREEDY-RC},
proving \cref{thm:TGREEDY-RC-2.5-approx}.

\section{Reducing Maximum Compression in SCS-RC to Scaffolding}
\label{sec:max-compression-scaffolding}

The overlap graph $G_{\mathrm{ov}}(S)$,
defined in \cref{sec:prefix-and-overlap-graphs},
is a complete weighted directed graph with vertex
set $V=S\cup\mathrm{rc}(S)$ and edge weights
$|\mathrm{ov}(u,v)|$ for $(u,v)\in V\times V$.
As discussed in \cref{sec:prefix-and-overlap-graphs},
an optimal SCS-RC solution corresponds to a
directed path visiting exactly one vertex from
$\{s,\mathrm{rc}(s)\}$ for each $s\in S$.
We call such a path an \emph{RC-admissible path}.
Thus, maximizing compression in SCS-RC is
equivalent to finding a maximum-weight
RC-admissible path in $G_{\mathrm{ov}}(S)$.
We reduce this problem to the scaffolding problem
defined in \cref{def:scaffolding-problem}.

Consider a complete weighted undirected graph
$G'=(V',E')$, where
$V'=\bigcup_{s\in S}\{s^+,s^-\}$ and
$E'=\binom{V'}{2}$.
The vertices $s^+$ and $s^-$ represent
$s$ and $\mathrm{rc}(s)$, respectively.
We treat them as distinct even when
$s=\mathrm{rc}(s)$, so that
$M=\{\{s^+,s^-\}\mid s\in S\}$
forms a perfect matching of $G'$.
For each vertex $u\in V'$, let $\lambda(u)$
denote its represented string.
We assign each edge $\{u,v\}\in E'$ the weight
\[
w(u,v)=
\begin{cases}
0 & \text{if } \{u,v\}\in M,\\
|\mathrm{ov}(\lambda(u),\mathrm{rc}(\lambda(v)))|
& \text{otherwise}.
\end{cases}
\]
This is well-defined since
$|\mathrm{ov}(x,\mathrm{rc}(y))|
=|\mathrm{ov}(y,\mathrm{rc}(x))|$
for any strings $x,y$.

We show that every RC-admissible path in
$G_{\mathrm{ov}}(S)$ can be transformed into
a Hamiltonian path in $G'$ containing $M$
with the same weight, and vice versa.

Every Hamiltonian path in $G'$ containing $M$
can be written as
$(u_1,\bar u_1,\ldots,u_n,\bar u_n)$,
where $\{u_i,\bar u_i\}\in M$ for each $i$.
Let $x_i=\lambda(\bar u_i)$.
Then $(x_1,\ldots,x_n)$ is an RC-admissible path,
and
\[
w(\bar u_i,u_{i+1})
=|\mathrm{ov}(x_i,x_{i+1})|.
\]
Since the matching edges have weight zero,
the two paths have the same weight.

Conversely, given any RC-admissible path
$(x_1,\ldots,x_n)$, choose $\bar u_i$ representing
$x_i$ and let $u_i$ be its matching partner.
Then $(u_1,\bar u_1,\ldots,u_n,\bar u_n)$
is a Hamiltonian path containing $M$,
and the same equality preserves its weight.

Therefore, this gives a weight-preserving reduction
from maximum compression in SCS-RC to scaffolding,
proving \cref{thm:scs-rc-max-compress-to-scaffolding}.
We use the following consequence of \cite{CHEN2018131}.

\begin{lemma}[{\cite{CHEN2018131}}] \label{lm:scaffolding-approx}
For every fixed $\varepsilon>0$, scaffolding
admits a polynomial-time
$(5/9-\varepsilon)$-approximation algorithm.
\end{lemma}
More precisely, \cite{CHEN2018131} establishes
the approximation ratio
$(5-5\varepsilon)/(9-8\varepsilon)$.
For simplicity, we use the weaker bound
$5/9-\varepsilon$.
Combining \cref{thm:MGREEDY-RC-3-approx,thm:scs-rc-max-compress-to-scaffolding} and \cref{lm:MGREEDY-RC-compress-framework,lm:scaffolding-approx},
we obtain a $(22/9+\varepsilon)$-approximation
for SCS-RC, proving \cref{thm:scs-rc-2.45-approx}.

\section*{Disclosure of AI Usage}
We used ChatGPT (GPT-6) to assist in developing proofs.
The manuscript was written by the authors,
with AI assistance in checking for potential
mathematical errors and polishing the English.
The authors have verified all results and
take full responsibility for the manuscript.



\bibliography{lipics-v2021-sample-article}

@article{blum1994linear,
author = {Blum, Avrim and Jiang, Tao and Li, Ming and Tromp, John and Yannakakis, Mihalis},
title = {Linear approximation of shortest superstrings},
year = {1994},
issue_date = {July 1994},
publisher = {Association for Computing Machinery},
address = {New York, NY, USA},
volume = {41},
number = {4},
issn = {0004-5411},
doi = {10.1145/179812.179818},
journal = {J. ACM},
month = jul,
pages = {630–647},
numpages = {18}
}

@article{JIANG1992195,
title = {A note on shortest superstrings with flipping},
journal = {Information Processing Letters},
volume = {44},
number = {4},
pages = {195-199},
year = {1992},
issn = {0020-0190},
doi = {https://doi.org/10.1016/0020-0190(92)90084-9},
author = {Tao Jiang and Ming Li and Ding-zhu Du},
}

@article{FICI2016245,
title = {On the greedy algorithm for the Shortest Common Superstring problem with reversals},
journal = {Information Processing Letters},
volume = {116},
number = {3},
pages = {245-251},
year = {2016},
issn = {0020-0190},
doi = {https://doi.org/10.1016/j.ipl.2015.11.015},
author = {Gabriele Fici and Tomasz Kociumaka and Jakub Radoszewski and Wojciech Rytter and Tomasz Waleń},
}

@InProceedings{yamano_et_al:LIPIcs.CPM.2026.15,
  author =	{Yamano, Ryosuke and Shibuya, Tetsuo},
  title =	{{Improved Approximation Ratios for the Shortest Common Superstring Problem with Reverse Complements}},
  booktitle =	{37th Annual Symposium on Combinatorial Pattern Matching (CPM 2026)},
  pages =	{15:1--15:11},
  series =	{Leibniz International Proceedings in Informatics (LIPIcs)},
  ISBN =	{978-3-95977-420-8},
  ISSN =	{1868-8969},
  year =	{2026},
  volume =	{369},
  editor =	{Bille, Philip and Prezza, Nicola},
  publisher =	{Schloss Dagstuhl -- Leibniz-Zentrum f{\"u}r Informatik},
  address =	{Dagstuhl, Germany},
  URN =		{urn:nbn:de:0030-drops-259412},
  doi =		{10.4230/LIPIcs.CPM.2026.15}
}

@article{lyndon1954burnside,
  author  = {Lyndon, Roger C.},
  title   = {On Burnside's Problem},
  journal = {Transactions of the American Mathematical Society},
  volume  = {77},
  number  = {2},
  pages   = {202--215},
  year    = {1954},
  doi     = {10.1090/S0002-9947-1954-0064049-X}
}

@techreport{chukhin2026tight,
  author      = {Nikolai Chukhin and Alexander S. Kulikov
                 and Ivan Mihajlin and Alexander Smal},
  title       = {A Tight Cycle-Cover Inequality for
                 Shortest Common Superstring},
  institution = {Electronic Colloquium on Computational Complexity
                 (ECCC)},
  number      = {TR26-157},
  year        = {2026},
  url         = {https://eccc.weizmann.ac.il/report/2026/157/}
}

@article{CHEN2018131,
title = {Approximation algorithms for the scaffolding problem and its generalizations},
journal = {Theoretical Computer Science},
volume = {734},
pages = {131-141},
year = {2018},
issn = {0304-3975},
doi = {https://doi.org/10.1016/j.tcs.2017.03.042},
author = {Zhi-Zhong Chen and Youta Harada and Fei Guo and Lusheng Wang},
}

@Inbook{Gevezes2014,
author="Gevezes, Theodoros P.
and Pitsoulis, Leonidas S.",
title="The Shortest Superstring Problem",
bookTitle="Optimization in Science and Engineering: In Honor of the 60th Birthday of Panos M. Pardalos",
year="2014",
publisher="Springer New York",
address="New York, NY",
pages="189--227",
isbn="978-1-4939-0808-0",
doi="10.1007/978-1-4939-0808-0_10",
}

@inproceedings{li1990towards,
author = {Li, M.},
title = {Towards a {DNA} sequencing theory (learning a string)},
year = {1990},
isbn = {081862082X},
publisher = {IEEE Computer Society},
address = {USA},
doi = {10.1109/FSCS.1990.89531},
booktitle = {Proceedings of the 31st Annual Symposium on Foundations of Computer Science},
pages = {125–134 vol.1},
series = {SFCS '90}
}

@article{KececiogluMyers1995,
  author  = {John D. Kececioglu and Eugene W. Myers},
  title   = {Combinatorial Algorithms for {DNA} Sequence Assembly},
  journal = {Algorithmica},
  volume  = {13},
  number  = {1-2},
  pages   = {7--51},
  year    = {1995},
  doi     = {10.1007/BF01188580}
}

@article{medvedev2009maximum,
author = {Paul Medvedev and Michael Brudno},
title ={Maximum Likelihood Genome Assembly},
journal = {Journal of Computational Biology},
volume = {16},
number = {8},
pages = {1101-1116},
year = {2009},
doi = {10.1089/cmb.2009.0047},
note ={PMID: 19645596},
}

@article{MyersJr+2016+126+132,
title = {A history of {DNA} sequence assembly},
author = {Eugene W. Myers Jr},
pages = {126--132},
volume = {58},
number = {3},
journal = {it - Information Technology},
doi = {10.1515/itit-2015-0047},
year = {2016},
lastchecked = {2026-11-09}
}

@INPROCEEDINGS{CazauxCanovasRivals2016,
  author={Cazaux, Bastien and Cánovas, Rodrigo and Rivals, Eric},
  booktitle={2016 Data Compression Conference (DCC)}, 
  title={Shortest {DNA} Cyclic Cover in Compressed Space}, 
  year={2016},
  volume={},
  number={},
  pages={536-545},
  doi={10.1109/DCC.2016.79}}

@article{conway2011succinct,
    author = {Conway, Thomas C. and Bromage, Andrew J.},
    title = {Succinct data structures for assembling large genomes},
    journal = {Bioinformatics},
    volume = {27},
    number = {4},
    pages = {479-486},
    year = {2011},
    month = {02},
    issn = {1367-4803},
    doi = {10.1093/bioinformatics/btq697},
}

@InProceedings{CPM1996.2+2/3approx,
author="Armen, Chris
and Stein, Clifford",
editor="Hirschberg, Dan
and Myers, Gene",
title="A 2 2/3-approximation algorithm for the shortest superstring problem",
booktitle="Combinatorial Pattern Matching",
year="1996",
publisher="Springer Berlin Heidelberg",
address="Berlin, Heidelberg",
pages="87--101",
isbn="978-3-540-68390-2"
}

@article{Breslaure.1997.OverlapRotationLemma,
author = {Breslauer, Dany and Jiang, Tao and Jiang, Zhigen},
title = {Rotations of Periodic Strings and Short Superstrings},
year = {1997},
issue_date = {Aug. 1997},
publisher = {Academic Press, Inc.},
address = {USA},
volume = {24},
number = {2},
issn = {0196-6774},
doi = {10.1006/jagm.1997.0861},
journal = {J. Algorithms},
month = aug,
pages = {340–353},
numpages = {14}
}

@article{SIAM.2+1/2approx,
author = {Sweedyk, Z.},
title = {A 2$\frac{1}{2}$-Approximation Algorithm for Shortest Superstring},
journal = {SIAM Journal on Computing},
volume = {29},
number = {3},
pages = {954-986},
year = {2000},
doi = {10.1137/S0097539796324661},
}

@article{KAPLAN200513.greedy3.5n,
title = {The greedy algorithm for shortest superstrings},
journal = {Information Processing Letters},
volume = {93},
number = {1},
pages = {13-17},
year = {2005},
issn = {0020-0190},
doi = {https://doi.org/10.1016/j.ipl.2004.09.012},
author = {Haim Kaplan and Nira Shafrir}
}

@inbook{SODA13.2+11/23approx,
author = {Marcin Mucha},
title = {Lyndon Words and Short Superstrings},
year = {2013},
publisher = {Proceedings of the 2013 Annual ACM-SIAM Symposium on Discrete Algorithms (SODA)},
pages = {958-972},
doi = {10.1137/1.9781611973105.69},
}

@inproceedings{STOC22.ImprovedApproximateGuarantees,
author = {Englert, Matthias and Matsakis, Nicolaos and Vesel\'{y}, Pavel},
title = {Improved approximation guarantees for shortest superstrings using cycle classification by overlap to length ratios},
year = {2022},
isbn = {9781450392648},
publisher = {Association for Computing Machinery},
address = {New York, NY, USA},
doi = {10.1145/3519935.3520001},
booktitle = {Proceedings of the 54th Annual ACM SIGACT Symposium on Theory of Computing},
pages = {317–330},
numpages = {14},
location = {Rome, Italy},
series = {STOC 2022}
}

@InProceedings{englert_et_al:LIPIcs.ISAAC.2023.29,
  author =	{Englert, Matthias and Matsakis, Nicolaos and Vesel\'{y}, Pavel},
  title =	{{Approximation Guarantees for Shortest Superstrings: Simpler and Better}},
  booktitle =	{34th International Symposium on Algorithms and Computation (ISAAC 2023)},
  pages =	{29:1--29:17},
  series =	{Leibniz International Proceedings in Informatics (LIPIcs)},
  ISBN =	{978-3-95977-289-1},
  ISSN =	{1868-8969},
  year =	{2023},
  volume =	{283},
  editor =	{Iwata, Satoru and Kakimura, Naonori},
  publisher =	{Schloss Dagstuhl -- Leibniz-Zentrum f{\"u}r Informatik},
  address =	{Dagstuhl, Germany},
  URN =		{urn:nbn:de:0030-drops-193319},
  doi =		{10.4230/LIPIcs.ISAAC.2023.29}
}

@misc{OAI:A-Polynomial-Time-2-Approximation-for-Shortest-Common-Superstring-September-24-2026,
  author = {{OpenAI}},
  title = {{A Polynomial-Time 2-Approximation for Shortest Common Superstring}},
  howpublished = {OpenAI Math Release preprint
                  \href{https://github.com/openai/math/blob/main/preprints/A-Polynomial-Time-2-Approximation-for-Shortest-Common-Superstring-September-24-2026/paper.pdf}{OAI:A-Polynomial-Time-2-Approximation-for-Shortest-Common-Superstring-September-24-2026}},
  year = {2026}
}

@article{Kaplan.MAXATSP.2/3approx,
author = {Kaplan, Haim and Lewenstein, Moshe and Shafrir, Nira and Sviridenko, Maxim},
title = {Approximation algorithms for asymmetric TSP by decomposing directed regular multigraphs},
year = {2005},
issue_date = {July 2005},
publisher = {Association for Computing Machinery},
address = {New York, NY, USA},
volume = {52},
number = {4},
issn = {0004-5411},
doi = {10.1145/1082036.1082041},
journal = {J. ACM},
month = jul,
pages = {602–626},
numpages = {25}
}

@InProceedings{paluch_et_al:LIPIcs.STACS.2012.501,
  author =	{Paluch, Katarzyna and Elbassioni, Khaled and van Zuylen, Anke},
  title =	{{Simpler Approximation of the Maximum Asymmetric Traveling Salesman Problem}},
  booktitle =	{29th International Symposium on Theoretical Aspects of Computer Science (STACS 2012)},
  pages =	{501--506},
  series =	{Leibniz International Proceedings in Informatics (LIPIcs)},
  ISBN =	{978-3-939897-35-4},
  ISSN =	{1868-8969},
  year =	{2012},
  volume =	{14},
  editor =	{D\"{u}rr, Christoph and Wilke, Thomas},
  publisher =	{Schloss Dagstuhl -- Leibniz-Zentrum f{\"u}r Informatik},
  address =	{Dagstuhl, Germany},
  URN =		{urn:nbn:de:0030-drops-34129},
  doi =		{10.4230/LIPIcs.STACS.2012.501}
}

@InProceedings{yamano_et_al:LIPIcs.WABI.2026.2,
  author =	{Yamano, Ryosuke and Shibuya, Tetsuo},
  title =	{{Improved Approximation Algorithms and Hardness Results for Shortest Common Superstring with Reverse Complements}},
  booktitle =	{26th International Conference on Algorithms for Bioinformatics (WABI 2026)},
  pages =	{2:1--2:13},
  series =	{Leibniz International Proceedings in Informatics (LIPIcs)},
  ISBN =	{978-3-95977-446-8},
  ISSN =	{1868-8969},
  year =	{2026},
  volume =	{390},
  editor =	{El-Mabrouk, Nadia and Vandin, Fabio},
  publisher =	{Schloss Dagstuhl -- Leibniz-Zentrum f{\"u}r Informatik},
  address =	{Dagstuhl, Germany},
  URN =		{urn:nbn:de:0030-drops-275069},
  doi =		{10.4230/LIPIcs.WABI.2026.2}
}

\appendix

\end{document}